\documentclass{optica-article}

\journal{opticajournal} 

\articletype{Research Article}
\usepackage{fancyhdr}
\usepackage{lineno}
\usepackage{bbm}
\usepackage{mathrsfs}
\usepackage{amsmath}

\begin{document}

\title{Flexible generation of optomagnonic quantum entanglement and quantum coherence difference in double-cavity-optomagnomechanical system}

\author{Xiaomin Liu,\authormark{1,2} Rongguo Yang,\authormark{1,2,3,4} Jing Zhang,\authormark{1,2,3,*} and Tiancai Zhang\authormark{1,2,3}}

\address{\authormark{1}College of Physics and Electronic Engineering, Shanxi University, Taiyuan 030006, China\\
\authormark{2}State Key Laboratory of Quantum Optics Technologies and Devices, Shanxi University, Taiyuan 030006, China\\
\authormark{3}Collaborative Innovation Center of Extreme Optics, Shanxi University, Taiyuan 030006, China\\
\authormark{4}yrg@sxu.edu.cn}
\email{\authormark{*}zjj@sxu.edu.cn}

\begin{abstract*}
Quantum entanglement and quantum coherence generated from the optomagnomechanical system are important resources in quantum information and quantum computation. In this paper, a scheme for flexibly generating optomagnonic quantum entanglement and quantum coherence difference is proposed, based on a double-cavity-optomagnomechanical system. The parameter dependencies of the bipartite optomagnonic entanglement, the genuine tripartite optomagnonic entanglement, the quantum coherence difference, and the stability of the system, are investigated intensively. The results show that this scheme endows the magnon more flexibility to choose different mechanisms, under the condition of maintaining the system stable. This work is valuable for connecting different nodes in quantum networks and manipulating the magnon states with light in the future.

\end{abstract*}

\section{Introduction}
As quantum information technology advances, quantum entanglement and quantum coherence, which are profound features of quantum theory, are essential resources for quantum computation\cite{45}, quantum communication\cite{965}, and quantum networks\cite{202266,2024579}, due to their functions of correlating nodes. The phonon and magnon modes in optomagnomechanical systems can interact with almost any physical mode at various frequency, which make the system a promising platform to generate entanglement and coherence, for connecting different components in hybrid quantum systems\cite{093601,1098,281,153,426,589,20171}. 
Therefore, abundant works have been done to discuss the generation of different hybrid entanglements based on optomagnomechanical systems\cite{015014,34764,s11128,2200866,202206}. An optomagnomechanical system composed of an optical cavity, and a microbridge-structured YIG crystal attached with a small high-reflectivity mirror pad, was proposed and studied, while the steady optomagnonic entanglement can be achieved when the optical cavity resonated with the anti-Stokes band of the driven laser field and the magnon mode resonated with the Stokes band of the driven microwave field\cite{015014}. If the cavity mode was driven by a blue-detuned laser field and the magnon mode was driven by a red-detuned microwave field, optomagnonic entanglement could also be obtained in a similar optomagnomechanical system\cite{202206}. Adding a microwave cavity in the above system, both optomagnonic and optomicrowave entanglements could be generated, when the magnon mode and microwave mode resonated with the Stokes band of driven microwave field and the optical mode resonated with the anti-Stokes band of driven laser field\cite{2200866}. If the adding microwave cavity was a gainful one, which made the system PT-symmetric-like, the generation and enhancement of optomagnonic and optomicrowave entanglements were also investigated, under the same condition\cite{34764}. Considering about the hybrid cavity optomagnomechanical system with double microwave cavities and double magnon modes, the microwave-microwave and magnon-magnon entanglements could be obtained when the microwave and magnon modes both resonated with the driven microwave field and the cavity mode resonated with the anti-Stokes band of driven laser field\cite{s11128}. 
On the other hand, the genuine tripartite entanglement, which involves stronger correlation and nonlocality\cite{86419}, plays a key role in some special quantum tasks, such as secure quantum key distribution protocols\cite{217903,499507}, quantum error correction\cite{602605}, etc. In recent years, the genuine tripartite optomagnonic entanglements generated from different optomagnomechanical systems were discussed\cite{202206,34764}.
In addition, as another kinds of quantum characters, quantum coherence and quantum coherence difference shown in various optomechanical and magnomechanical systems have attracted more attentions\cite{052314,215502,11204}. Especially, the quantum coherence difference, which is related with the mutual information, also describes the correlation between different modes. The total quantum coherence and the coherence difference of the cavity and mechanical modes in a cavity optomechanical system were achieved when the cavity mode was driven by a red-detuned laser\cite{052314}. Replacing two cavity mirrors with Laguerre–Gaussian rotating mirror and adding a two-level atom ensemble in the above system, the quantum optomechanical coherence could be confirmed and enhanced if the atoms were far-off-resonant with the driven field\cite{215502}. In a hybrid Laguerre–Gaussian cavity magnomechanical system including two Yttrium iron garnet (YIG) spheres, quantum coherence of two YIG spheres could be obtained when the two YIG spheres resonated with the driven microwave field and the microwave cavity resonated with the anti-Stokes band of the driven microwave field\cite{11204}. 
Note that all of the above-mentioned works require certain mechanism of each mode, to create macroscopic quantum effects, such as the bipartite and tripartite entanglements and the quantum coherence. In this paper, a flexible scheme of generating bipartite and genuine tripartite optomagnonic entanglements and quantum coherence difference, based on a double-cavity-optomagnomechanical system, are investigated in detail. Here the magnon mode has more space to choose a blue-detuned or red-detuned driving microwave field, that is, it is allowed to be manipulated by adjusting the frequency of the driven microwave field, under the condition of keeping stability of the system. 

\section{Generation of bipartite and genuine tripartite optomagnonic entanglements and quantum coherence difference}
\begin{figure}[htbp]
\centering\includegraphics[width=13cm]{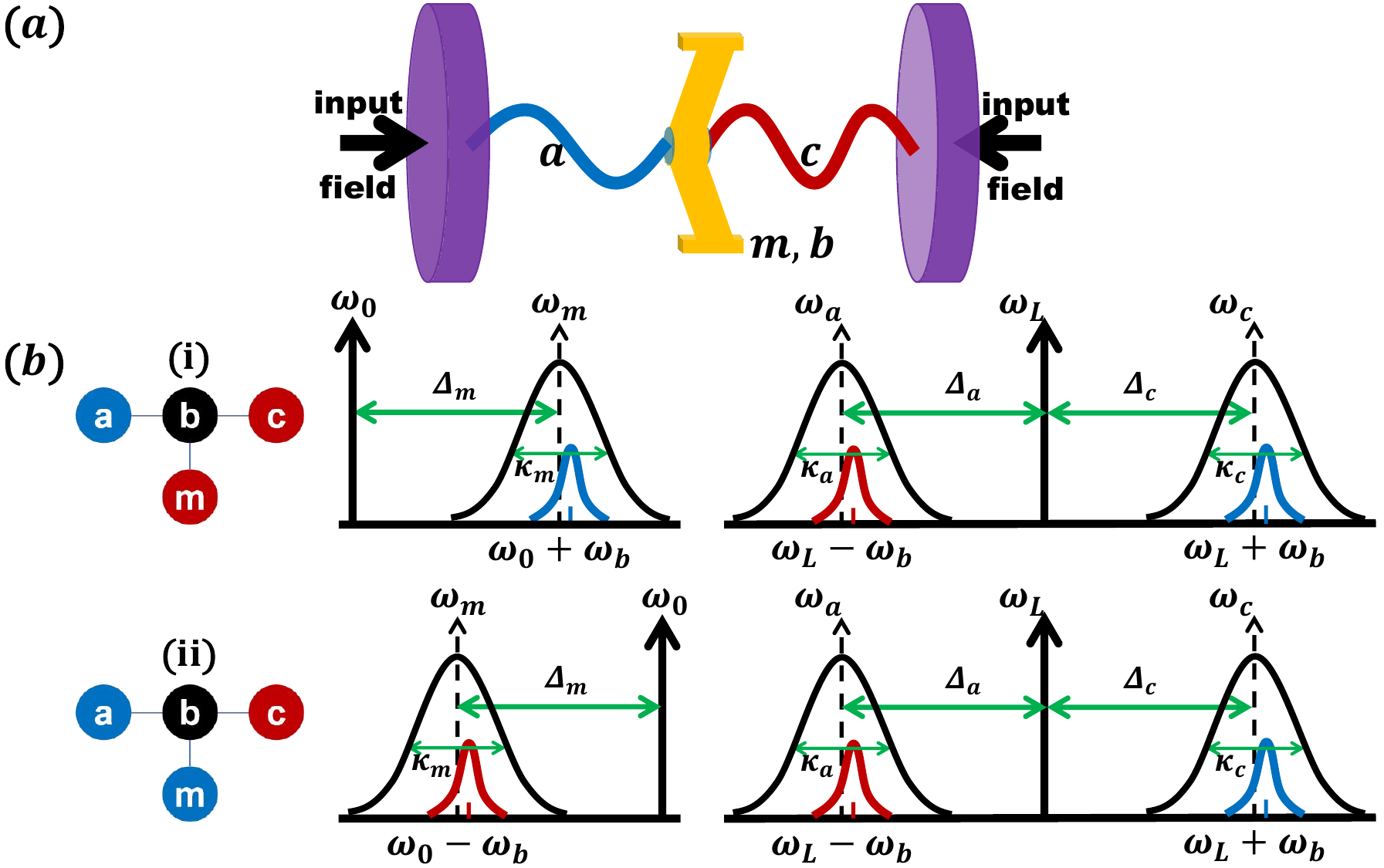}
\caption{The scheme of generating flexible quantum optomagnonic entanglement and coherence difference in system. (a) Schematic diagram. (b) frequency relation.}
\end{figure}
The considered optomagnomechanical system is shown in Fig.1(a). The system includes two optical cavities (with frequency $\omega_a,\omega_c$ and decay $\kappa_a,\kappa_c$) composed of two mirrors and a microbridge-structured YIG crystal (with frequency $\omega_m(\omega_b)$ and decay $\kappa_m (\gamma_b)$ for its magnon (phonon) mode), both sides of which are attached with two high-reflectivity mirrors. The two cavity modes are driven by laser beams with frequency $\omega_L$, and the YIG crystal is driven by a microwave field with frequency $\omega_0$. Considering the condition that the cavity modes $a$ and $c$ resonate with the Stokes (red) and anti-Stokes (blue) sidebands of the driven laser beam, respectively, as shown in Fig.1(b). On the one hand, the blue-detuned driven optical cavity mode $a$ can be used to create entanglement, through a down-conversion process; on the other hand, the red-detuned driven optical cavity $c$ can be used to transfer the entanglement and cool the lower-frequency mechanical motion, through a state-swap process\cite{1391,30005,29581}. Therefore, the magnon mode $m$ will have more flexibility, i.e. it can be driven by either a red-detuned one (to transfer entanglement, scheme (i)) or a blue-detuned microwave field (to create entanglement, scheme (ii))\cite{031201,0370,070101}, as shown in Fig.1(b). The Hamiltonian of this system can be described as:
{\begin{equation}
 \begin{aligned}
 \hat{H}/\hbar &= \omega_a \hat{a}^{\dagger} \hat{a}+\omega_{c} \hat{c}^{\dagger} \hat{c}+\omega_{m} \hat{m}^{\dagger} \hat{m}+\frac{\omega_b}2(\hat{q}^2+\hat{p}^2) \quad \\
 &\quad - g_{a} \hat{a}^{\dagger}\hat{a}\hat{q}+ g_{c} \hat{c}^{\dagger}\hat{c}\hat{q}+ g_{m} \hat{m}^{\dagger}\hat{m}\hat{q} \quad \\
 &\quad + i\eta_a(\hat{a}^{\dagger}e^{-i\omega_Lt}- \hat{a}e^{i\omega_Lt})+i\eta_c(\hat{c}^{\dagger} e^{-i\omega_{L} t}- \hat{c} e^{i\omega_{L} t})\quad \\
 &\quad +i\Omega(\hat{m}^{\dagger} e^{-i\omega_{0} t}- \hat{m} e^{i\omega_{0} t}),
 \end{aligned}
\end{equation} }
where $\hat{o}^{\dagger}$ ($\hat{o}$) is the creation (annihilation) operator of the mode $o$ with frequency $\omega_{o}$ and decay rate $\kappa_{o}$, $o=a, c, m$. $\hat{p}$ and $\hat{q}$ are the dimensionless momentum and position operators of the phonon mode, respectively. $\eta_j=\sqrt{2P\kappa_j/\hbar\omega_j}$, $j=a,c$, denotes the cavity-laser coupling strengths, $\Omega=\frac{\sqrt{5}}{4}\gamma\sqrt{N_d}B_d$ denotes the Rabi frequency related to the microwave drive (with frequency $\omega_0$ and magnetic component $B_d$)\cite{203601}, where $P$ is the input power, $\gamma$ is the gyromagnetic ratio and $N_d$ is the total number of spins of YIG bridge. The four terms of the first row of Eq.(1) are the free Hamiltonians of the cavity modes ($a, c$), the magnon mode $m$, and the phonon mode $b$, respectively. The three terms of the second row of Eq.(1) represent the optomechanical and magnomechanical couplings. The third and fourth rows are the driving terms of the cavity modes and the magnon mode, respectively.
From the Hamiltonian described by Eq.(1), the linearized quantum Langevin equations of the quadrature fluctuations ($\delta X_a$, $\delta Y_a$, $\delta X_{c}$, $\delta Y_{c}$, $\delta X_{m}$, $\delta Y_{m}$, $\delta q$, $\delta p$ ) can be obtained as $\dot{u}=\mathcal{A} u + n(t)$ (details are shown in Appendix), where $\delta X_{o}=(\delta o+\delta o^{\dagger})/\sqrt{2}$, and $\delta Y_{o}=i(\delta o^{\dagger}-\delta o)/\sqrt{2}, o=a, c, m$. $\Delta_{o}$ is the effective frequency detuning of the mode $o$. Then the corresponding covariance matrix ($CM$) can be obtained by solving the Lyapunov equation. Logarithmic negativity $E_N$\cite{032314,090503}, minimum residual contangle $R^{min}$\cite{7821,200615} and quantum coherence difference $\Delta C_{Q}$\cite{20070370,621,L21,065102} are adopted to quantificate bipartite optomagnonic entanglement, genuine tripartite optomagnonic entanglement and corresponding optomagnonic intercoherence, respectively.
In this paper, except for the variables, the relative parameters are: $\omega_b/2\pi=25MHz$, $\kappa_a/2\pi=1MHz$, $\kappa_c/2\pi=2MHz$, $\kappa_m/2\pi=1MHz$, $\gamma_b/2\pi=100Hz$, $T=10mK$, $G_a/2\pi=1.5MHz$, $G_c/2\pi=8MHz$, $\Delta_a=-\omega_b$, $\Delta_c=\omega_b$. For scheme (i) $\Delta_m=\omega_b$, $G_m/2\pi=6MHz$, and for scheme (ii) $\Delta_m=-\omega_b$, $G_m/2\pi=1.5MHz$.

\begin{figure}[htbp]
\centering\includegraphics[width=13.2cm]{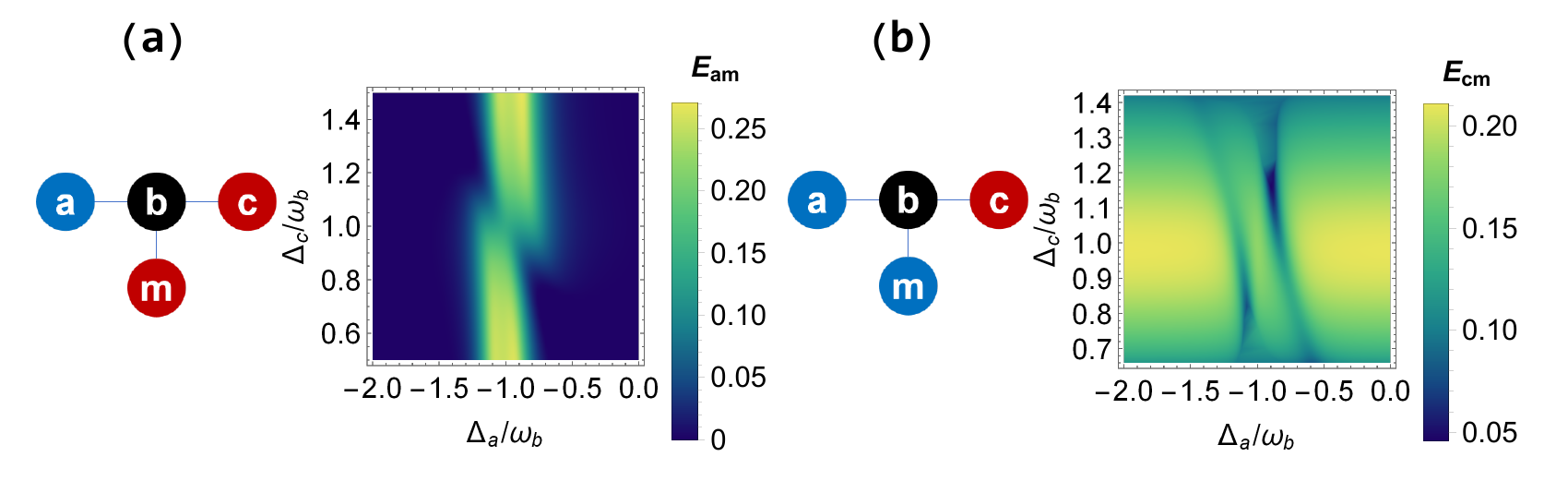}
\caption{The optomagnonic entanglement $E_{am}$ and $E_{cm}$ versus detunings $\Delta_{a}$ and $\Delta_{c}$, where $E_{am}$ and $E_{cm}$ are generated from scheme (i) and (ii), respectively. In schematic diagrams, blue and red circles represent the blue and red detuned driven, respectively.}
\end{figure}
The obtained optomagnonic entanglements $E_{am}$ and $E_{cm}$, which are generated from scheme (i) and (ii), respectively, vary with the detunings ($\Delta_a$ and $\Delta_c$) are shown in Fig.2(a) and Fig.2(b). The entanglement $E_{am}$ becomes strongest near $\Delta_a=-\omega_b$, because it is the optimal parameter condition of generating optomechanical entanglement between the cavity mode $a$ and the phonon mode $b$, and the entanglement can be further distributed to the magnon mode $m$ through magnomechanical coupling between the magnon mode $m$ and the phonon mode $b$, resulting in a stationary entanglement of $E_{am}$, as seen in Fig.2(a).
While the maximum $E_{cm}$ occurs mainly near $\Delta_c=\omega_b$, because the entanglement between the magnon mode $m$ and the phonon mode $b$ is the best here, which can be further distributed to the cavity mode $c$ through optomechanical coupling between the cavity mode $c$ and the phonon mode $b$, yielding a stationary entanglement of $E_{cm}$, as seen in Fig.2(b).
In addition, the anti-crossing around $\Delta_a=-\omega_b$ is a signature of the strong coupling. $E_{am}$ and $E_{cm}$ perform reversely due to the different requirements of magnomechanical coupling (state transfer dominated for $E_{am}$, down-conversion dominated for $E_{cm}$).

\begin{figure}[htbp]
\centering\includegraphics[width=13cm]{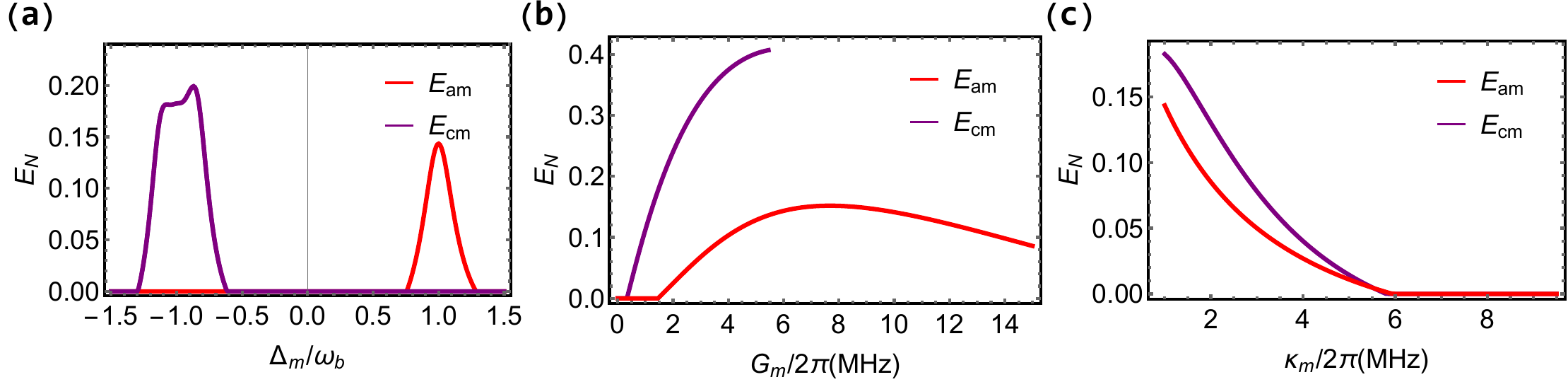}
\caption{The optomagnonic entanglement $E_{am}$ (red) and $E_{cm}$ (purple) versus $\Delta_m$(a), $G_{m}$(b) and $\kappa_{m}$(c), respectively.}
\end{figure}
First of all, optomagnonic entanglements $E_{am}$ and $E_{cm}$ versus the magnon-related parameters are shown in Fig.3. 
It is clear that $E_{am}$ and $E_{cm}$ are mainly around $\Delta_m=\omega_b$ (red detuned) and $\Delta_m=-\omega_b$ (blue detuned), respectively, as shown in Fig.3(a). Because optomagnonic entanglements $E_{am}$ and $E_{cm}$ come mainly from the mechanisms of transferring and creating entanglements, respectively, for the magnon mode $m$.
In Fig.3(b), when the magnomechanical coupling strength $G_m$ is increased, the optomagnonic entanglement $E_{am}$ will first increase and then decrease. The essence of this phenomenon is the competition between the cavity mode $a$ and the magnon mode $m$ for the phonon mode $b$. This can be further understood by the Bogoliubov transformation: $\alpha_1=a \cosh r+m^\dagger \sinh r$, $\alpha_2=m \cosh r+a^\dagger \sinh r$, where, $r=arc\tanh\frac{G_m}{G_a}$. Then, the coupling strength between the Bogoliubov modes and the phonon is $G=\sqrt{G_a^2-G_m^2}$. However, stronger $r$ requires weaker $G_a$ and stronger $G_m$, while stronger $G$ requires stronger $G_a$ and weaker $G_m$, leading to the non-monotonic behavior of $E_{am}$. For the entanglement $E_{cm}$, it increases with the increase of $G_m$, due to the increase of the created entanglement between the magnon mode $m$ and the phonon mode $b$. Note that the system becomes unstable if $G_m$ is further increased, which will be discussed later. It can be seen that $E_{cm}$ is stronger than $E_{am}$, because the relative strength of the magnomechanical coupling in scheme (ii) is stronger than in scheme (i). 
As shown in Fig.3(c), $E_{am}$ and $E_{cm}$ decrease with increasing magnon decay $\kappa_m$, due to the decrease in the generation and transfer of magnon-involved entanglement. 
\begin{figure}[htbp]
\centering\includegraphics[width=13cm]{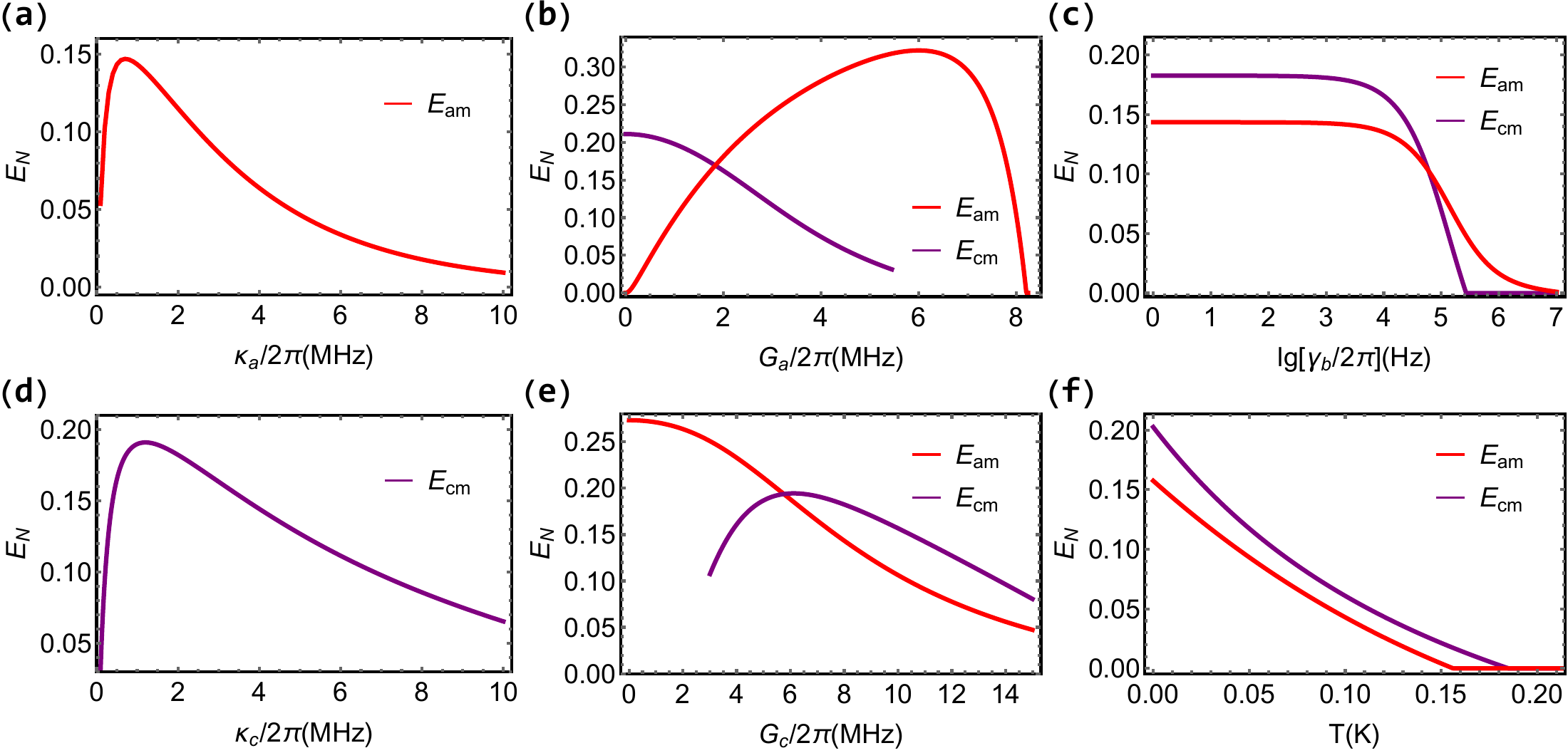}
\caption{The optomagnonic entanglements $E_{am}$ (red) and $E_{cm}$ (purple) versus $\kappa_a$(a), $\kappa_c$(d), $G_a$(b), $G_c$(e), $\gamma_b$(c) and $T$(f), respectively.}
\end{figure}
Moreover, how the optomagnonic entanglements $E_{am}$ and $E_{cm}$ vary with other parameters can be found in Fig.4.
As shown in Fig.4(a) and Fig.4(d), optomagnonic entanglements $E_{am}$ and $E_{cm}$ first increase dramatically, then decrease slowly, versus cavity decay rates $\kappa_a$ and $\kappa_c$, respectively. It can be indicated that, although dissipation is usually regarded as a negative factor of entanglement generation, small dissipation is positive for generating entanglement through dissipation coupling. 
In Fig.4(b), on the one hand, with increasing optomechanical coupling strength $G_a$, $E_{am}$ increases in a wide range, because stronger optomechanical coupling $G_a$ helps generate stronger optomechanical entanglement between cavity mode $a$ and phonon mode $b$, which also contributes to optomagnonic entanglement $E_{am}$. However, when $G_a$ is too strong, the entanglement transferred to magnon mode $m$ will be less, which is negative for the optomagnonic entanglement $E_{am}$. On the other hand, with increasing optomechanical coupling strength $G_a$, $E_{cm}$ decreases during the period of $E_{am}$ increasing, under our parameter conditions. This can be understood by the competition between the cavity mode $a$ and the magnon mode $m$ for phonon $b$.
The behaviors of optomagnonic entanglements $E_{cm}$ and $E_{am}$ in Fig.4(e) can be explained in the same way. 
The quality of phonon is also a factor that should be considered in generating optomagnonic entanglement, because it is the intermediate to connect other modes. It can be seen from Fig.4(c) that the optomagnonic entanglements $E_{am}$ and $E_{cm}$ are almost unchanged until the decay rate increases to $\gamma_b/2\pi\thicksim 10^4Hz$ and then decrease with further increasing $\gamma_b$. Note that $E_{am}$ is more robust than $E_{cm}$ due to the stronger cooling mechanism.
Quantum entanglement is usually very fragile, so the ability of the system to resist environmental disturbance is of great concern. The entanglements decrease with increasing temperature $T$ and can survive above $150mK$ under our parameters, as shown in Fig.4(f), which means that the optomagnonic entanglements $E_{am}$ and $E_{cm}$ are relatively robust to thermal noise. 

\begin{figure}[htbp]
\centering\includegraphics[width=13.2cm]{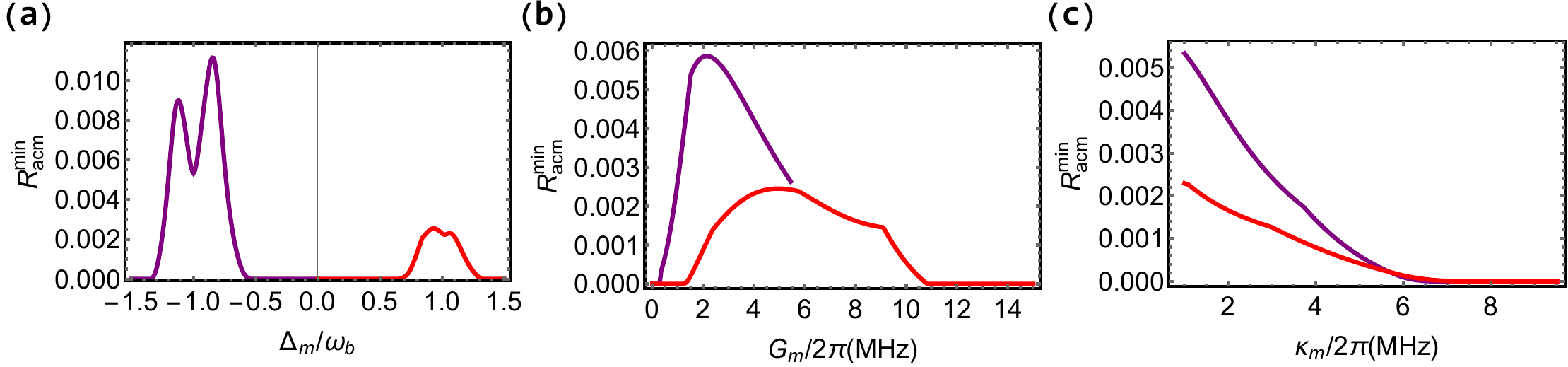}
\caption{The genuine tripartite entanglement $R_{acm}^{min}$ versus magnon-related parameters $\Delta_m$(a), $G_{m}$(b) and $\kappa_{m}$(c) in two schemes (red and purple curves correspond to the scheme (i) and (ii), respectively).}
\end{figure}
What is more, the genuine tripartite entanglement among modes $a$, $c$ and $m$, which is indicated by the non-zero minimum residual contangle $R^{min}_{acm}$, can be observed in schemes (i) and (ii), as shown in red and purple in Fig.5, respectively. Compared with bipartite entanglements in Fig.3(a), the genuine tripartite entanglements in Fig.5(a) show more obvious signatures of the strong coupling. In Fig.5(b), similar to bipartite optomagnonic entanglements, genuine tripartite entanglements obtained in schemes (i) and (ii) both show a trend of "first increase and then decrease" with increasing magnomechanical coupling strength $G_m$, due to the competition among cavity mode $a$, cavity mode $c$ and magnon mode $m$ for phonon mode $b$. In Fig.5(c), genuine tripartite entanglements both decrease with increasing $\kappa_m$, due to the weaker interaction involving the magnon mode $m$.
\begin{figure}[htbp]
\centering\includegraphics[width=13.2cm]{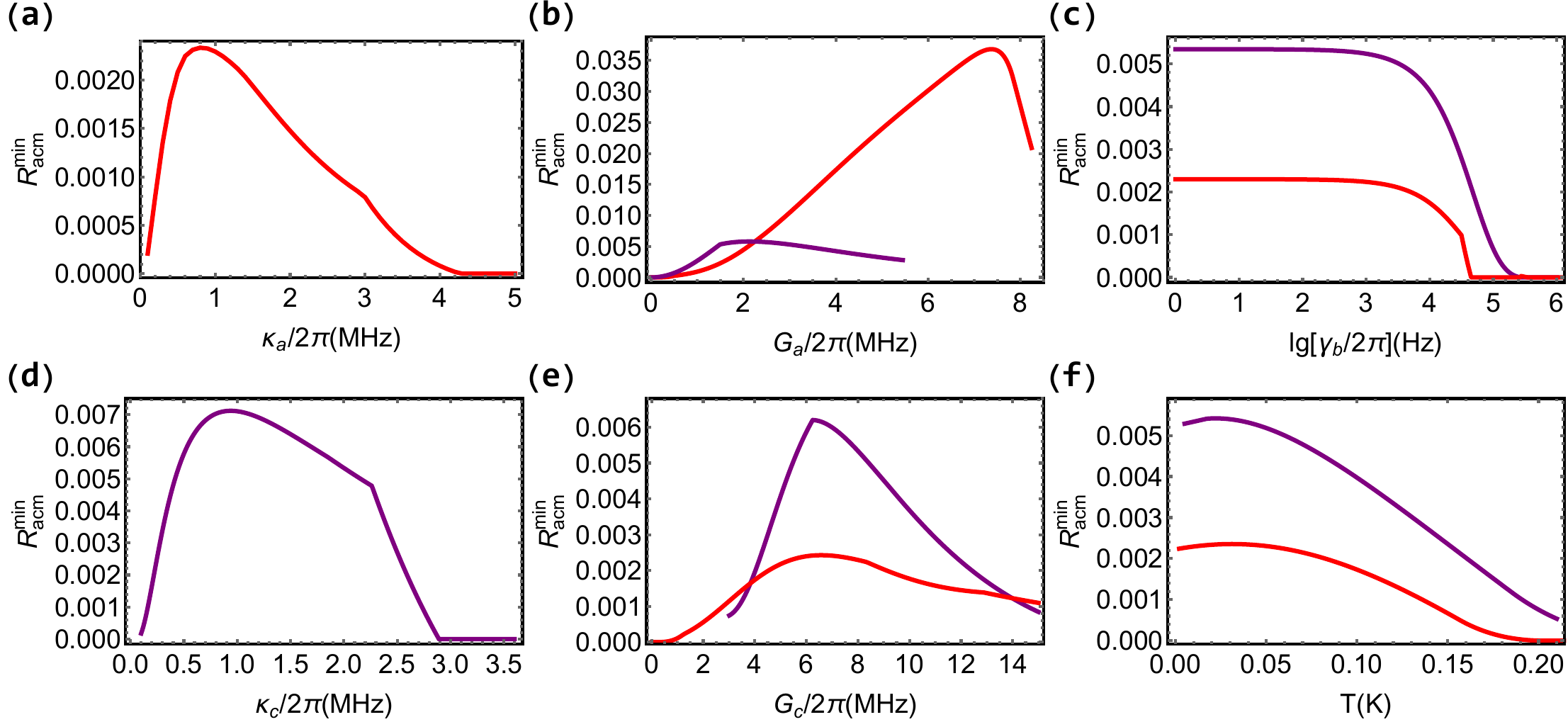}
\caption{The genuine tripartite entanglement $R_{acm}^{min}$ versus parameters $\kappa_a$(a), $\kappa_c$(d), $G_a$(b), $G_c$(e), $\gamma_b$(c) and $T$(f) in two schemes (red and purple curves correspond to the scheme (i) and (ii), respectively).}
\end{figure}
The genuine tripartite entanglement can also be observed in schemes (i) and (ii), as shown in red and purple in Fig.6, respectively. The genuine tripartite entanglements versus cavity decay rates $\kappa_a$ and $\kappa_c$ are shown in Fig.6(a) and Fig.6(d), respectively, which is similar to that of the bipartite entanglement. It can be seen that genuine tripartite entanglements can exist only at small cavity decay rate, indicating that genuine tripartite entanglements are more sensitive than bipartite entanglements to cavity decay rates. In Fig.6(b) and Fig.6(e), genuine tripartite entanglements show a similar non-monotonous trend to bipartite entanglements with increasing optomechanical coupling strengths $G_a$ and $G_c$. In Fig.6(c), genuine tripartite entanglements present a similar monotonous trend to bipartite entanglements with increasing phonon decay rate $\gamma_b$, because the interaction ability of the phonon mode is compromised by larger phonon decay. In Fig.6(f), genuine tripartite entanglements exhibit stronger robustness against temperature changes compared to bipartite entanglements, due to the weaker affection from the average number of thermal excitations for both the magnon mode and the phonon mode. Overall, the curves of genuine tripartite entanglements versus different parameters are not so smooth compared to those of bipartite entanglements, due to the competition among more modes for phonon mode.

\begin{figure}[htbp]
\centering\includegraphics[width=13cm]{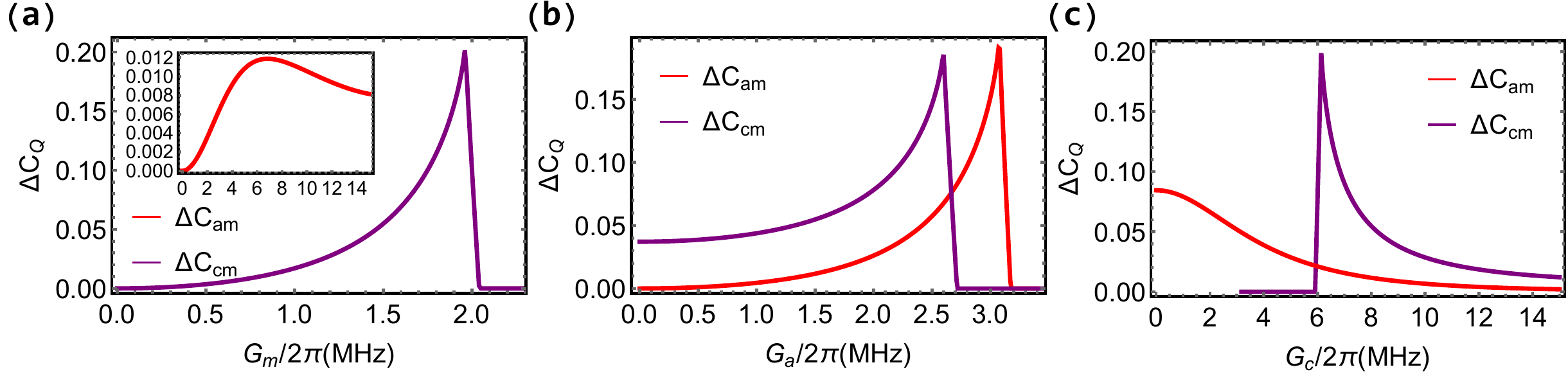}
\caption{The quantum coherence difference $\Delta C_{am}$ (red) and $\Delta C_{cm}$ (purple) versus magnon coupling $G_m$(a), cavity mode $a$ coupling $G_a$(b) and cavity mode $c$ coupling $G_c$(c), respectively.}
\end{figure}
Furthermore, the quantum coherence difference describes the intercoherence between two modes and quantifies how much one mode correlates with the other mode. The quantum coherence differences versus different parameters are discussed in Fig.7 and Fig.8. Similarly, the quantum coherence differences $\Delta C_{am}$ and $\Delta C_{cm}$ can also be observed in schemes (i) and (ii), as shown in red and purple, respectively. In Fig.7(a) and Fig.7(b), $\Delta C_{cm}$ increases with increasing magnomechanical and optomechanical coupling strengths $G_m$ and $G_a$, while $\Delta C_{am}$ increases with increasing $G_a$ and various non-monotonously (first increases, then decreases) with increasing $G_m$. In Fig.7(c), the quantum coherence differences decrease with increasing the optomechanical coupling strength $G_c$. It seems that the quantum coherent differences are sensitive to the mechanism at work, and it can exist in a narrow coupling strength region under the entanglement creation mechanism and wider coupling strength region under the state transfer mechanism. This suggests that, when the entanglement creation mechanism dominates, stronger couplings help to strengthen the optomagnonic intercoherence. In contrast, when the state transfer mechanism dominates, stronger couplings will weaken the optomagnonic intercoherence. Note that the "sharply down to zero" behavior of intercoherence means that the system is going to be unstable.
\begin{figure}[htbp]
\centering\includegraphics[width=13cm]{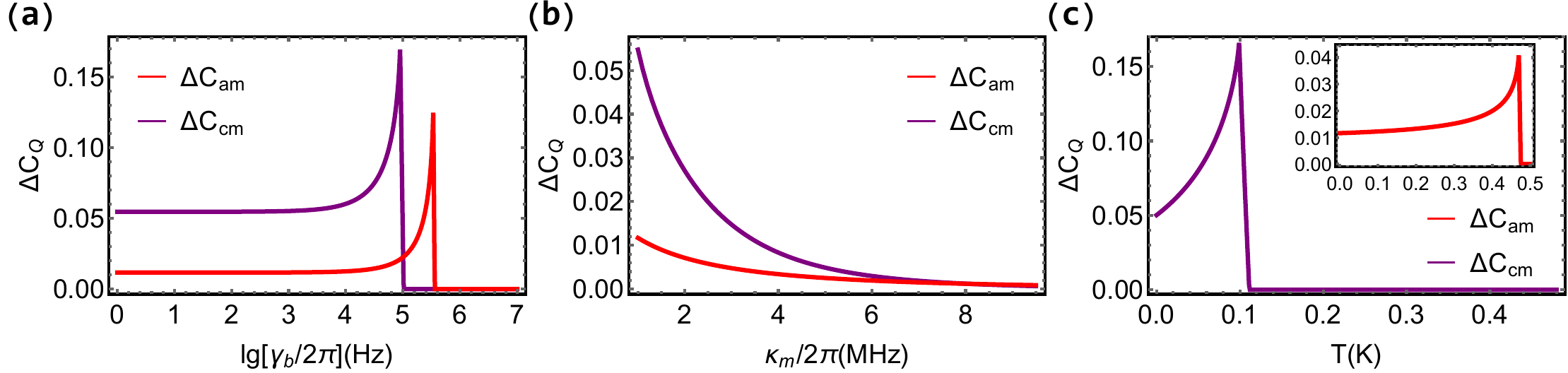}
\caption{The quantum coherence difference $\Delta C_{am}$ (red) and $\Delta C_{cm}$ (purple) versus phonon decay $\gamma_{b}$(a), magnon decay $\kappa_{m}$(b) and temperature $T$(c), respectively.}
\end{figure}
The effect of decay channels on optomagnonic intercoherence is shown in Fig.8. It can be found in Fig.8(a) that the quantum coherence differences are almost unchanged until the decay rate increases to $\gamma_b/2\pi\thicksim 10^4Hz$, and after that increase rapidly and then sharply down to zero with further increasing $\gamma_b$ due to the entanglement creation mechanism. The parameter regions in which $\Delta C_{am}$ and $\Delta C_{cm}$ can exist stably are wide enough, with the former being wider than the latter, due to the stronger state transfer mechanism. In Fig.8(b), the quantum coherence differences decrease with increasing $\kappa_m$, due to weaker interaction mechanism of magnon. In Fig.8(c), with increasing temperature $T$, both quantum coherence differences first increase, which is because that more modes have better resistance to thermal noise than single mode, then sharply down to zero. Meanwhile, $\Delta C_{am}$ can exist in a wider range of $T$ than $\Delta C_{cm}$, due to the stronger state transfer mechanism.

\section{Analysis of system's stability}
\begin{figure}[htbp]
\centering\includegraphics[width=10cm]{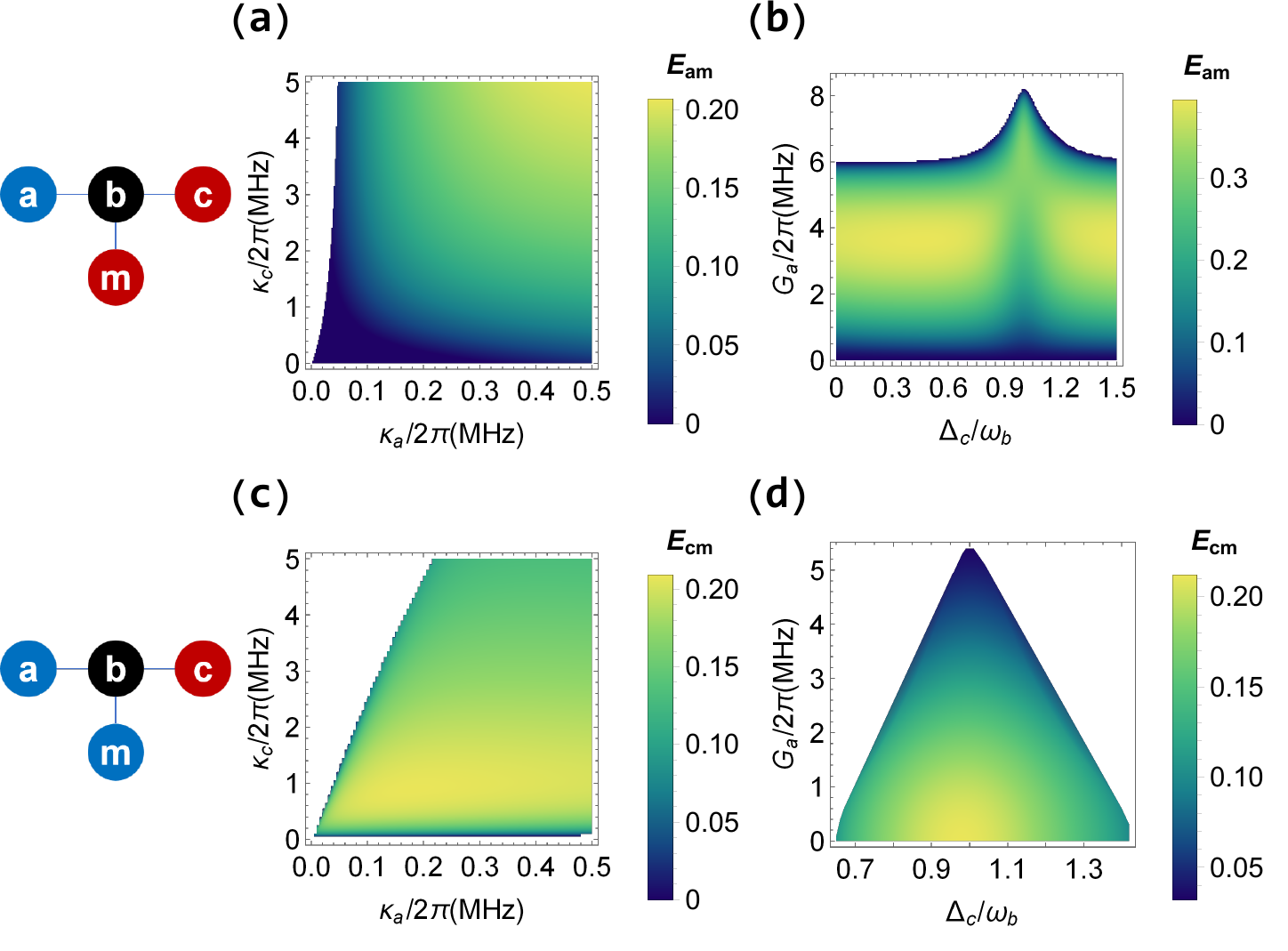}
\caption{The optomagnonic entanglements $E_{am}$ and $E_{cm}$ versus $\kappa_{a}$(a,c) and $\Delta_c$(b,d) in two schemes, respectively.}
\end{figure} 
Quantum entanglement and quantum coherence exist only within the stable region of the system, judged by the negativity of the real parts of all the eigenvalues of the drift matrix $\mathcal{A}$. Therefore, it is necessary to discuss the stable region of the system. The different parameter regions in which stability exists are shown in Fig.9 and Fig.10.
The stable regions depending on cavity decays for schemes (i) and (ii) are shown in Fig.9(a) and Fig.9(c), respectively. Smaller $\kappa_a$ and larger $\kappa_c$ will lead to greater instability of the system, corresponding to weaker state transfer and stronger entanglement creation abilities, respectively. However, under common experimental conditions, the stability of the system can be guaranteed.
The stable regions depending on $G_a$ and $\Delta_c$ for schemes (i) and (ii) are shown in Fig.9(b) and Fig.9(d), respectively. When cavity mode $c$ resonates with the anti-Stokes sideband of the driven field, the system tends to be stable. In contrast, if it gradually deviates from the anti-Stokes sideband, the system will be more and more unstable. Stronger coupling $G_a$ and larger distance from the anti-Stokes sideband will make the system more unstable, which corresponds to stronger entanglement creation and weaker state transfer abilities, respectively. 
\begin{figure}[htbp]
\centering\includegraphics[width=10cm]{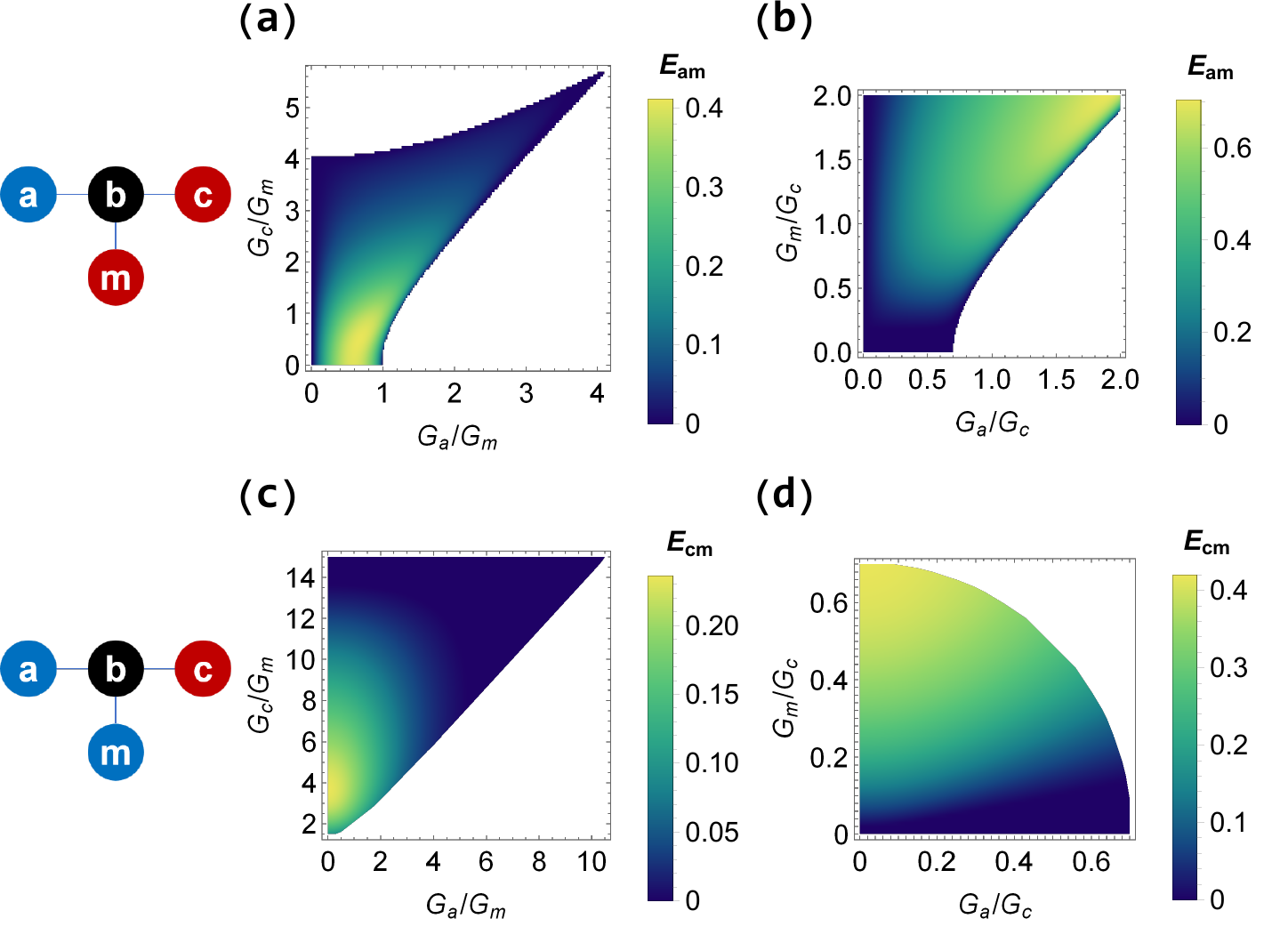}
\caption{The optomagnonic entanglements $E_{am}$ and $E_{cm}$ versus $G_a/G_m$(a,c) and $G_a/G_c$(b,d) in two schemes, respectively.}
\end{figure}
The stable regions depending on the coupling strengths $G_a$ and $G_c$ (normalized to $G_m$) for schemes (i) and (ii) are shown in Fig.10(a) and Fig.10(c), respectively.
It is found that smaller $G_c$ and larger $G_a$ will lead to larger instability of the system, corresponding to weaker state transfer and stronger entanglement creation abilities, respectively. However, too large $G_c$ may also make the system unstable, due to too strong state transfer ability, as shown in Fig.10(a).
Moreover, the stable regions depending on the coupling strengths $G_a$ and $G_m$ (normalized to $G_c$) for schemes (i) and (ii) are shown in Fig.10(b) and Fig.10(d), respectively. When the magnon mode $m$ is driven by a red-detuned microwave field, smaller $G_m$ and larger $G_a$ will lead to larger instability of the system, corresponding to weaker state transfer and stronger entanglement creation abilities, respectively, as shown in Fig.10(b). In contrast, for a blue-detuned-driven magnon, larger $G_a$ and $G_m$ will lead to instability of the system, corresponding to stronger entanglement creation abilities, as shown in Fig.10(d).
Overall, it can be found that the area of the stability region is related to the strength and type of mechanism at work, that is, weak entanglement-creation and mediate state-transfer mechanisms help the system to keep stable. In addition, by comparing all these figures, it is obvious that the areas of the stable regions in scheme (i) are larger than in scheme (ii), due to the stronger state transfer ability (the phonon can be cooled effectively); however, the state transfer ability is not allowed to be too strong, because it starts to play a role of generating entanglement\cite{030405}.

\section{Conclusion}
In this paper, we propose a scheme for flexibly generating optomagnonic quantum entanglement and quantum coherence difference, based on a double-cavity-optomagnomechanical system. We intensively investigate the parameter dependence of the bipartite optomagnonic entanglement, the genuine tripartite optomagnonic entanglement, and the quantum coherence difference, to obtain the optimal parameter condition. Moreover, we also study the area of the stable parameter region and analyze the parameter condition in which the system could stay stable. It is interesting that the magnon mode can be driven either by a red-detuned microwave field (scheme (i)) or by a blue-detuned microwave field (scheme (ii)), which makes a more flexible choice of the magnon mode. In other words, one can freely switch between scheme (i) and (ii), by adjusting the frequency of the driven microwave field. It is helpful to understand the inetraction mechanism in optomagnomechanical systems and to consider about the effective manipulation of the various kinds of quantum correlation in macroscopic quantum systems. It also provides a theoretical basis for controlling, designing, detecting, and transmitting magnon's states with light, allowing greater flexibility in manipulating and utilizing the quantum properties of magnon.

\section{Appendix}
Here we provide the details on how we obtain the steady-state entanglement. The entanglement is calculated based on the covariance matrix among the cavity modes, the magnon mode and the phonon mode. The covariance matrix can be achieved by solving the linearized quantum Langevin equations, which can be rewritten in the following form:
\begin{equation}
\dot{u}(t)=\mathcal{A} u(t)+n(t)
\end{equation}
where $u(t)$ is the vector of quadrature fluctuation operators of cavity modes, magnon mode and phonon mode. $\mathcal{A}$ is the drift matrix and $n(t)$ is the vector of noise quadrature operators associated with the noise terms. 

The steady-state covariance matrix $V(t\rightarrow\infty)$ of the system quadratures, with its entries defined as $V_{ij} = \frac{1}{2} \langle{\{u_i(t),u_j(t)\}}\rangle$, which can be obtained by solving the Lyapunov equation:
\begin{equation}
\mathcal{A}V+V\mathcal{A}^{T}=-\mathcal{D}
\end{equation}
where the diffusion matrix $\mathcal{D}$ is defined as
$\frac{1}{2} \langle{n_i(t) n_j(t^{\prime}) +n_j(t^{\prime}) n_i(t)}\rangle=\mathcal{D}_{ij}\delta(t-t^{\prime})$. $a_{in}(t)$, $c_{in}(t)$, $m_{in}(t)$ and $\xi(t)$ are the optical noise operators of the cavity modes $a,c$, the magnon mode $m$ and the phonon mode $b$, respectively, which are zero mean and characterized by the following correlation function: 
{\begin{equation}\begin{aligned}
\left \langle a_{in}(t)a_{in}^{\dagger}(t^{\prime}) \right \rangle&=\delta(t-t^{\prime})\\\left \langle c_{in}(t)c_{in}^{\dagger}(t^{\prime}) \right \rangle&=\delta(t-t^{\prime})\\\left \langle m_{in}(t)m_{in}^{\dagger}(t^{\prime}) \right \rangle&=({\overline{n}_m+1})\delta(t-t^{\prime})\\\left \langle m_{in}^{\dagger}(t)m_{in}(t^{\prime}) \right \rangle&={\overline{n}_m}\delta(t-t^{\prime})\\\left \langle \xi(t)\xi(t^{\prime})+\xi(t^{\prime}) \xi(t)\right \rangle/2 &\simeq\gamma_{b}(2{\overline{n}_b}+1)\delta(t-t^{\prime}) 
\end{aligned}\end{equation}}
where a Markovian approximation has been made, and ${\overline{n}_m}=\left[exp[(\hbar \omega_{m}/k_{B}T)]-1\right]^{-1}$ and ${\overline{n}_b}=\left[exp[(\hbar \omega_{b}/k_{B}T)]-1\right]^{-1}$ are the equilibrium mean magnon and phonon, respectively.

Once the covariance matrix $V$ is obtained, the entanglement can then be quantified by means of logarithmic negativity:
\begin{equation}
E_N=max[0, -ln2\nu_{-}]
\end{equation}
where $\nu_{-} = \min \thinspace eig\lvert i\Omega_2V_m\rvert$ ($\Omega_2= \oplus_{j=1}^2 i \sigma_y $ is the so-called symplectic matrix and $\sigma_y$ is the $y$ Pauli matrix) is the minimum symplectic eigenvalue of the covariance matrix $V_m= PVP$, with $V_m$ being the $4 \times 4$ covariance matrix associated of interest, and $P = diag\begin{pmatrix} 1, 1, 1, -1 \end{pmatrix}$ the matrix that inverts the sign of momentum, $p_2\rightarrow -p_2$, realizing partial transposition at the level of covariance matrices.

A $bona$ $fide$ quantification of tripartite entanglement is given by the minimum residual contangle:
\begin{equation}
R^{min}=min[R_{i|jk},R_{j|ik},R_{k|ij}]
\end{equation}
where $R_{i|jk}=C_{i|jk}-C_{i|j}-C_{i|k}\geq 0$ is the residual contangle, with $C_{u|v}$ the contangle of subsystems of u and v (v contains one or two modes), which is a proper entanglement monotone defined as the squared logarithmic negativity. A nonzero minimum residual $R^{min}>0$ denotes the presence of genuine tripartite entanglement in the system.

Then, we investigate the coherence of a single-mode Gaussian state. For a given bosonic mode $\hat{o}$ with the commutation relation $[\hat{o},\hat{o}^\dagger] = 1$, we can define the quadrature operators $\hat{X}_o=(\hat{o}+\hat{o}^\dagger)/\sqrt{2}$, $\hat{Y}_o=i(\hat{o}^\dagger-\hat{o})/\sqrt{2}$. In general, for a Gaussian state $\rho$, it can be fully described by its first moment $\vec{o}=(\vec{o}_1,\vec{o}_2)=Tr(\rho \hat{\vec{o}})$ (steady-state solution) and second moment (steady-state covariance matrix). The coherence of any given one-mode Gaussian state $\rho(V,\vec{o})$ is defined as
\begin{equation}
 C_Q(\rho(V,\vec{o}))=-F(\nu)+F(2\overline{n}+1)
\end{equation}
where $\nu=\sqrt{Det(V)}$ is symplectic eigenvalue of covariance matrix $V$, $\overline{n}=\frac{V_{11}+V_{22}+\vec{o}_1^2+\vec{o}_2^2-2}{4}$, $F(x)=\frac{x+1}{2}\log_2(\frac{x+1}{2})-\frac{x-1}{2}\log_2(\frac{x-1}{2})$.
For a two-mode Gaussian state system, the quantum coherence difference $\Delta C_Q$ can be expressed as
\begin{equation}
\begin{aligned}
 \Delta C_Q&=C_{Q}-C_{Q_1}-C_{Q_2}\\ &=F(\nu_1)+F(\nu_2)-F(\nu_{-})-F(\nu_{+})
\end{aligned}
\end{equation}
where $\nu_{+}$ and $\nu_{-}$ are two symplectic eigenvalues of the covariance matrix.

In the frame rotating at the driven fields with frequency $\omega_L$ and $\omega_0$, the quantum Langevin equations ($QLEs$) describing the system, can be written as:
{\begin{equation}\begin{aligned}
\dot{a} & = -(i\Delta_{a}^0+\kappa_a)a +ig_{a}aq +\eta_a +\sqrt{2\kappa_a}a_{in}(t)\\
\dot{c} & = -(i\Delta_{c}^0+\kappa_c)c -ig_{c}cq +\eta_c +\sqrt{2\kappa_c}c_{in}(t)\\
\dot{m} & = -(i\Delta_{m}^0+\kappa_m)m -ig_{m}mq +\Omega +\sqrt{2\kappa_m}m_{in}(t)\\
\dot{q} & = \omega_b p\\
\dot{p} & = -\omega_b q-\gamma_b p +g_{a}a^\dagger a -g_{c} c^\dagger c -g_{m} m^\dagger m +\xi(t),
\end{aligned}\end{equation}}
The linearized QLEs can also be expressed as the form of Eq.(6) with\\
$\mathcal{A}=\begin{pmatrix}
 -\kappa_a & \Delta_{a} & 0 & 0 & 0 & 0 & G_{a} & 0\\
 -\Delta_{a} &-\kappa_a & 0 & 0 & 0 & 0 & 0 & 0\\
 0 & 0 & -\kappa_c & \Delta_{c} & 0 & 0& -G_{c} & 0\\
 0 & 0 &-\Delta_{c} &-\kappa_c & 0 & 0 & 0 & 0\\
 0 & 0 & 0 & 0 & -\kappa_m & \Delta_{m} & -G_{m} & 0\\
 0 & 0 & 0 & 0 & -\Delta_{m} & -\kappa_m & 0 & 0\\
 0 & 0 & 0 & 0 & 0 & 0 & 0 & \omega_{b}\\
 0 & -G_{a} & 0 & G_{c} & 0 & G_{m} & -\omega_{b} & -\gamma_{b} \\
\end{pmatrix}$,\\
where $\Delta_{a}=\Delta_{a}^0-g_a q_{s}$ and $\Delta_{c(m)}=\Delta_{c(m)}^0+g_{c(m)} q_{s}$, $\Delta_{o}^0=\omega_{o}-\omega_{L(0)} (o=a, c, m)$ is the frequency detuning of the mode $o$ with respect to the corresponding driven fields in the frame rotating at the frequency of corresponding driven fields $\omega_L$ and $\omega_0$, respectively. $G_{o}= \sqrt{2}g_o o_s$, $o=a,c,m$. Here $a_s, c_s, m_s, q_{s}$ are the steady state solutions of the two cavity modes $a, c$, the magnon mode $m$ and the phonon mode $b$ ($a_s=\frac{\eta_a}{\kappa_a+i\Delta_a}$, $c_s=\frac{\eta_c}{\kappa_c+i\Delta_c}$, $m_s=\frac{\Omega}{\kappa_m+i\Delta_m}$, $q_s=\frac{-g_a a_s +g_c c_s +g_m m_s}{\omega_b}$) and $g_o$ is the single-photon coupling strength between the mode $o$ and the phonon mode. 

The corresponding matrix $\mathcal{D}$ of the Lyapunov equation Eq.(3) is $\mathcal{D}=Diag(\kappa_a,\kappa_a,\kappa_c,\kappa_c,\\
\kappa_m(2{\overline{n}_m}+1),\kappa_m(2{\overline{n}_m}+1),0,\gamma_{b}(2{\overline{n}_b}+1))$.

\begin{backmatter}
\bmsection{Funding}
This work was supported by: Innovation Program for Quantum Science and Technology (2023ZD0300400); National Key Research and Development Program of China (Grants No. 2021YFC2201802,  2021YFA1402002); and National Key Laboratory of Radar Signal Processing (JKW202401).
\bmsection{Disclosures}
The authors declare no conflicts of interest.
\bmsection{Data availability} Data underlying the results presented in this paper are not publicly available at this time but may be obtained from the authors upon reasonable request.
\end{backmatter}

\bibliography{ref}

\end{document}